\documentclass[aps,prd,10pt,twocolumn,showpacs]{revtex4-2}
\usepackage{amsmath,amssymb,amsfonts,dcolumn,color,graphicx,graphics,latexsym,placeins,epsfig}
\usepackage{epsfig}
\usepackage{bm}
\usepackage{latexsym}
\usepackage{natbib}
\usepackage{url}
\usepackage{dcolumn}
\usepackage{color}
\usepackage{amsfonts,amssymb,amsmath}
\usepackage{graphicx,epsfig}
\usepackage{psfrag}
\usepackage{subfigure}
\usepackage{hyperref}
\hypersetup{colorlinks=true}

\newcommand{\be}{\begin{equation}}
\newcommand{\ee}{\end{equation}}
\newcommand{\ba}{\begin{eqnarray}}
\newcommand{\ea}{\end{eqnarray}}

\begin{document}

\title{Constraints on higher curvature gravity from time delay between GW170817 and GRB 170817A}

\author{Avirup Ghosh$^{1}$}
\email{avirup.ghosh@iitgn.ac.in}
\author{Soumya Jana$^{2}$}
\email{sjana@prl.res.in}
\author{Akash K Mishra$^{1}$}
\email{akash.mishra@iitgn.ac.in}
\author{Sudipta Sarkar$^{1}$}
\email{sudiptas@iitgn.ac.in}

\affiliation{${}^{1}$ {\it  Indian Institute  of  Technology,  Gandhinagar 382355, India }}
\affiliation{${}^{2}${\it Physical Research Laboratory, Ahmedabad 380009, India }}

\begin{abstract}
General relativity may be an effective theory with higher curvature correction terms in the action. Inclusion of these terms leads to exciting new possibilities, e.g., gravitational and electromagnetic perturbations following different geodesics, leading to a time delay. Such a time delay was observed between the gravitational wave event GW170817 and its electromagnetic counterpart GRB 170817A. We describe how this effect can be used to constrain the coupling of the higher curvature term. Our method is sufficiently general and applicable to any higher curvature theory. 
\end{abstract}


\maketitle

\section{ Introduction}
 
 The recent detection of gravitational waves (GWs) by the LIGO Collaboration \cite{ligovirgo,gw170817} provides an unprecedented opportunity to test the theories of gravity beyond general relativity (GR). So far, no significant deviation from GR has been found in the weak-field regime, through several precision tests \cite{Will2014}. But, the gravitational wave astronomy allows us to test GR at the cosmological scales \cite{Ishak:2018his} as well as in the strong gravity regime. There are already interesting model-independent constraints on deviations from GR based on the observation of GW signals from compact black hole binaries \cite{grtest}.

Among the observed GW signals at LIGO-Virgo, GW170817 \cite{gw170817} is the only binary neutron star merger event with an electromagnetic counterpart, the gamma-ray burst  GRB 170817A \cite{grb170817a}. It opened up the window for multimessenger astronomy, leading to exciting physics such as constraining the theories beyond GR \cite{gwspeed,jana,gwfr} as well as probing the presence of extra dimensions \cite{ed}. The source was localized at a luminosity distance of $40$ Mpc at redshift $z\sim 0.0099$. Interestingly, the electromagnetic (EM) signal was observed $\sim 1.7$ s after the GW signal. The time delay can be explained as the difference in speeds of GW and EM signals constraining it to less than $10^{-15}c$ \cite{grb170817a}. Using this bound,  constraints on several theories beyond GR were also obtained \cite{gwspeed,jana}. The possibility of explaining such a time delay by gravitational lensing was discussed in \cite{lensing}. 

There are several motivations to seek physics beyond general relativity. The classical theory is perturbatively nonrenormalizable and therefore may make sense only as an effective theory, with new higher curvature correction terms in the low-energy effective action \cite{grplus}. The higher curvature gravity is the simplest possible modification of GR, and there is already a vast literature on the aspects of higher curvature gravity \cite{grplus_app}; some theoretical constraints on the higher curvature couplings are also known \cite{theory_bound}. Tests based on the modification of Newton's law at short length scales and other astrophysical tests from compact objects led to several stringent bounds \cite{obs_bound}. The gravitational wave observations provide another critical window to study the effect of higher curvature gravity at the cosmological scales.

 In this paper, we provide a general formalism to constrain higher curvature gravity using the observed time delay. In a generic higher curvature gravity, the graviton and photon follow different {\it paths} (geodesic) while propagating on a curved background \cite{characteristics}. This phenomenon may lead to a delay between gravitational and electromagnetic radiation. We study this effect and find an expression for the time delay. We discuss how assumptions related to the intrinsic delay at the event can influence the constraints. Our method is sufficiently general so that it can be used to study the time delay in any higher curvature modification of general relativity.\\

\section{Time delay between GW and EM signals: General setup }

We assume the homogeneous and isotropic FLRW line element as the background spacetime metric. The electromagnetic signal travels along the null geodesics of the metric,
\begin{equation}
ds^2=-dt^2+a^2(t)\left(dr^2+r^2d\Omega^2\right).
\label{eq:frw1}
\end{equation}
The background spacetime is a solution to the field equations of the underlying gravitational theory. The basic causal properties of such a theory are described by the characteristic hypersurfaces of the field equations. In GR, a hypersurface is characteristic if, and only if, it is null. As a result, the gravitational wave also follows the null geodesics of the metric given by Eq.~(\ref{eq:frw1}).  However, in higher curvature gravity, the study of characteristics of the perturbation equation shows that the gravitational radiation follows the null geodesic of an `effective metric' instead of the actual physical metric in Eq.~(\ref{eq:frw1}) \cite{characteristics}. The effective metric for GW is of the form
\begin{equation}
ds^2_{\rm eff}= -U(t)dt^2+ a^2(t)V(t)\left(dr^2+r^2d\Omega^2\right),
\label{eq:eff_metric}
\end{equation}     
where $U(t)$ and $V(t)$ come from the higher curvature terms. In the GR limit $U,V\rightarrow 1$. 

Suppose the EM signal is emitted from the source at the time $t_E$ and is observed at $t_O$. Let the time delay between the observation of GW and EM signal be  $\delta t_O$ and let $\delta t_E$ be the intrinsic delay in the emission of the GW at the source. It is expected that such an intrinsic delay must be present due to the difference in emission processes of GW and EM radiations. The extent of the intrinsic delay will play a significant role in constraining the physics beyond general relativity. In our convention, if the graviton arrives earlier then $\delta t_O$ is negative, while if it is emitted later then $\delta t_E$ is positive. Using the null geodesics of the background and the effective metric, one arrives at the following expression relating the time delays: 
\begin{eqnarray}\label{eq:td1}
\left(\frac{1}{a}\sqrt{\frac{U}{V}}\right)\Bigg|_{t_O}\, \delta t_O &-& \left(\frac{1}{a}\sqrt{\frac{U}{V}}\right)\Bigg|_{t_E}\, \delta t_E \nonumber\\&& = \int_{t_E}^{t_O}\frac{1}{a}\left(1- \sqrt{\frac{U}{V}}\right) dt.\label{general_exp}
\end{eqnarray}
We set the scale factor to be one at the present epoch, $a(t_O)=1$ and $a(t_E)=(1+z_E)^{-1}$,  where $z_E$ is the redshift of the merger event. Transforming the time integral in Eq.~(\ref{eq:td1}) into a redshift integral, we get
\begin{eqnarray}
\delta t_O &=& \delta t_E \, (1+z_E)\sqrt{\frac{U_EV_O}{U_OV_E}}\nonumber \\ 
&& +\sqrt{\frac{V_O}{U_O}}\int^{z_E}_0 \frac{dz}{H(z)}\left(1-\sqrt{\frac{U(z)}{V(z)}}\right),
\label{eq:td2}
\end{eqnarray}
where we use the notations $U(t_E)=U_E$, $V(t_E)=V_E$, $U(t_O)=U_O$, and $V(t_O)=V_O$. The Hubble parameter $H(z)$ is defined as $\dot{a}(t)/ a(t) $ expressed as a function of the redshift $z$. It is easy to check that in the GR limit, the observed time delay is equal to the redshifted intrinsic delay.
 
Equation (\ref{eq:td2}) is our main expression which will be used to constrain the modification of GR. We need to calculate the functions $U(z)$ and $V(z)$ for a specific theory and compare with the observations. We consider the theory described by the Lagrangian 
\begin{equation}
{\cal L} = R + a\, R^2 + b\, R_{ab}R^{ab} + c\, R_{abcd} R^{abcd}.\label{lagrag}
\end{equation}
In four dimensions, using the Gauss-Bonnet theorem, we can relate the last term with the previous two terms. Also, a pure $\text{Ricci} \, \text{scalar}^2$ term does not change the causal structure of the theory. As a result, we only need to consider the following theory:
\begin{equation}
{\cal L} = R +\alpha\, R_{ab}\,R^{ab} .\label{lagra}
\end{equation}
The higher curvature coupling $\alpha$ has the dimension of $\text{length}^2$. Any higher curvature theory, such as above has many pathological features. For example, in four dimensions, every higher curvature theory suffers from perturbative ghosts \cite{ghost} (see \cite{Salvio:2019ewf} for a new perspective). The initial value formalism may not be well defined. As a result, we will treat the higher curvature term only as the first correction term of an effective theory. Therefore, we will expand everything till the first order in $\alpha$, neglecting the higher-order contributions. The time delay will be determined up to the first order in $\alpha$, neglecting ${\cal O}(\alpha^2)$ terms, and it will then be compared with the observational result. It will also be evident that the same procedure can be repeated for any theory of gravity that is a small deviation from GR.\\

\section{Time delay in higher curvature gravity}

Next, we present the analysis of time delay in the context of higher curvature theory described by the Lagrangian in Eq. (\ref{lagra}). In this theory, the components of the effective graviton metric are \cite{geff_frw}\\
\begin{eqnarray}
&&U = \frac{1}{1+\alpha\, H^2\left(-3(1+z)\frac{HH^\prime}{H^2}+5\right)},\nonumber\\
&&V = \frac{1}{1+\alpha\, H^2\left(-(1+z)\frac{HH^\prime}{H^2}+5\right)}\label{effective metric},
\end{eqnarray}
where everything is written as functions of the redshift $z$ and the prime here denotes derivative with respect to $z$. To obtain the time delay, we use the effective graviton metric coefficients from Eq.~(\ref{effective metric}) and expand it as a power series in $\alpha$. At each order in $\alpha$, the coefficients are functions, which can be obtained as perturbative solutions to the Friedmann equation of the theory in Eq. (\ref{lagra}). At this point, it is necessary to point out certain technical details of the perturbative expansion. Since the redshift $z$ is an observable, we will take it as our variable to express various quantities. Hence, we rewrite the Friedmann equation completely in terms of $z$ and then find its solution order by order in $\alpha$. We will assume that the Hubble parameter $H(z)$ has the following expansion:
\begin{eqnarray}
H(z)&=& H_{G}(z)+\alpha \, h(z) +\mathcal{O}(\alpha^2).
\end{eqnarray} 
The zeroth-order solution $H_G(z)$ is nothing but the solution obtained for GR. The first-order perturbation, $h(z)$ can be determined by solving the field equations of the theory in Eq. (\ref{lagra}). The boundary condition for such a solution can be chosen such that the correction $h(z)$ is equal to zero at $z=0$. This is equivalent to the assumption that the theory we are looking at, today, is predominantly GR and that the higher curvature effects are dominant at high redshifts.  Most importantly, this guarantees that the present day density parameter for dark energy ($\Omega_\Lambda$) is equal to $0.7$ with that of matter ($\Omega_m$) taken to be $0.3$. Such a boundary condition is only a particular choice. But, since we are only interested in results up to the linear order in $\alpha$, we will not require the explicit form of $h(z)$. The final ${\cal O}(\alpha)$ result can be expressed by using the GR solution $H_G(z)$ only. We need not solve the actual Friedmann equations of the higher curvature theory. \\

The dimensionless small parameter in our expansion is $\eta = \alpha \, H^2_G (0)$ in natural units. In GR, we have $\eta = 0$ and any nonzero value of $\eta$ measures the contribution of the higher curvature coupling $\alpha$ in terms of the characteristic size of the background universe. We hope to constrain this by the time delay observation.  

 The intrinsic delay may also depend on the higher curvature coupling. Assuming that $\delta t_O$ and $\delta t_E$ have series expansions in terms of $\alpha$ and equating terms of the same order, obtained from Eq. (\ref{eq:td2}), one arrives at the following expressions:

\begin{eqnarray}
&&\text{Order} ~\eta^{(0)}:~~\delta t_O^{(0)} = \delta t_E^{(0)}(1+z_E)\nonumber\\
&&\text{Order} ~\eta^{(1)}:\nonumber\\
&&~~ \delta t_O^{(1)} = -\int_{0}^{z_E}(1+z)H^{\prime}_{\rm G}~dz\nonumber + \delta t_E^{(1)}(1+z_E)\nonumber\\
&&~~~~+\delta t_E^{(0)}(1+z_E)\bigg(H_G^{\prime}(z_E)H_G(z_E)-H_G^{\prime}(0)H_G(0)\bigg),\nonumber\\
\label{alphaexpansion}
\end{eqnarray}
where the numbers in superscripts imply perturbation order. The ${\cal O}(\eta^0)$ equation is the GR case when there is no contribution from the higher curvature terms, and the observed delay is equal to the redshifted value of the intrinsic delay. The ${\cal O}(\eta)$ equation gives the first-order correction to the GR result. As we mentioned before, the net observed delay in the observation of the EM signal can come from two sources. The first one is purely astrophysical and depends on the detailed mechanism of gamma ray bursts (GRBs).  The other is either from a modification of the theory of gravity, lensing, or Shapiro delay. To discuss the first, one must note that the emission mechanism for GRBs is not completely understood. There are several models, out of which the relativistic fireball model is the most accepted one. In this model, a fraction of the gravitational energy released during the merger is assumed to be utilized to form a fireball constituted of $e^{\pm}$, gamma rays, and baryons. The fireball must also expand relativistically with a high Lorentz factor ($\Gamma$), with respect to the central engine, to avoid depletion due to $\gamma\gamma$ interactions \cite{Meszaros:2006rc, Kumar:2014upa}. Due to this, the emission of gamma rays can occur from a position away from the central object. This distance appropriately converted to time, corrected by the Lorentz factor $\Gamma$, can attribute to the time delay \cite{Shoemaker:2017nqv}. If the outflow is in the form of a narrow jet, then this is further affected by the angle between the line of sight and the jet, as well as by the opening angle of the jet. There can also be a time offset between the emission of gravitational waves and the ejection of the outflow itself \cite{Shoemaker:2017nqv}.

Due to the absence of an independent estimate of the delay due to the astrophysical effects, we have assumed that it can be completely accounted for, by a term like $\delta t_E$ in Eq. (\ref{DDNDOA}) and concentrate more on the part of the delay arising from the modification of the gravity theory. Nevertheless, the intrinsic delay term cannot be completely segregated from the delay arising due to the modification of gravity. This is because the intrinsic delay itself undergoes a redshift that depends on the gravity theory. To obtain an initial estimate, we will assume that the intrinsic delay does not dependent on the higher curvature coupling $\alpha$. This will be the case if the astrophysical effects discussed above are independent of the underlying theory of gravity. Then, we have a simpler equation for the observed time delay,

\begin{eqnarray}\label{DDNDOA}
\delta t_O &=& \delta t_E (1+z_E) \left[1+ \alpha \left(H_G^{\prime} (z_E)H_{G}(z_E) \right. \right. \nonumber \\ && \left. \left.-H_G^{\prime}(0) H_{G}(0)\right)\right] -\alpha \int_{0}^{z_E}(1+z)H^{\prime}_{\rm G}dz.
\end{eqnarray}

Given the intrinsic delay $\delta t_E$ at the source, the above equation can be used to determine the quantity $\eta$. On substituting $H_{G}(z)=H_{G}(0)\sqrt{\Omega_m(1+z)^3+\Omega_\Lambda}$ and putting the various parameters for the merger event, we obtain, 

\begin{eqnarray}\label{DDNDOA}
\eta = \frac{ \delta t_O - 1.0099 \,\delta t_E}{0.00904275\, \delta t_E - 0.00451145\, t_H},\label{etasource}
\end{eqnarray}

where the quantity $t_H = 1/H_{G}(0)$. If there is no prior knowledge or estimate of the intrinsic time delay, we can only get an estimate of $\eta$ for various trial values of $\delta t_E$. In particular, let us first assume $\delta t_E = 0$, which will give us an estimate of the upper bound for $\eta$.  Then, using the appropriate factors for the speed of light $c$, one obtains

\begin{eqnarray}\label{etalimit}
\eta = \frac{\alpha   \, H^2_G (0)}{c^2} \leq 8.5 \times 10^{-16}.
\end{eqnarray}

This upper bound on $\eta$ translates into an upper bound on the higher curvature coupling as $\alpha \leq 10^{\,36} \, m^2  $ which is obviously a weak bound. The other tests have more stringent bounds on various models of higher curvature gravity \cite{obs_bound}. A similar weak bound was also obtained from the bound on the GW speed for another type of alternative gravity, possessing a nonlinear matter-gravity coupling instead of having higher derivative terms \cite{jana}. \\

Our result is important due to following reasons: This is a bound from cosmological considerations, which constrains the coupling $\alpha$ compared to the scale of the Universe, whereas most of the other constraints are from local tests. For example, the Newtonian limit of higher curvature terms in the Lagrangian introduces an extra Yukawa-like term in the gravitational potential. The E\"ot-Wash experiment tries to verify such a Yukawa-like additional term by measuring departures from the Newtonian potential. In fact, the  $R_{ab}R^{ab}$ theory also introduces a similar additional term \cite{Capozziello:2009ss}.  Such an experiment puts an upper bound $\lesssim 2\times 10^{-9}~\text{m}^2$ on the parameter in the Yukawa term \cite{Berry:2011pb}. But, this analysis cannot uniquely bound the $R_{ab} R^{ab}$ theory, as the departure from the Newtonian potential will be contributed by all possible higher curvature interactions of the form given in Eq. (\ref{lagrag}). Our result provides the possibility of having a bound on the coupling of  $R_{ab} R^{ab}$ term only. If we use the above bound from the E\"ot-Wash experiment in our case, we will have $\delta t_O - \delta t_E \sim 10^{-52}\text{s} $, which implies that the observed time delay is entirely due to the redshifted intrinsic delay. Similarly, Planetary precession rates put the upper bound $\lesssim 1.2\times 10^{18} \text{m}^2$ on the coupling of the $\text{Ricci} \, \text{scalar} ^ 2$ theory. If we assume a similar bound on $\alpha$ coming from such considerations, we will have $\delta t_O - \delta t_E \sim 10^{-19}\text{s} $. We again emphasize that these bounds are obtained for a pure $f(R)$ theory of gravity. We are using these bounds simply to obtain a rough estimate of the intrinsic time delay. As far as our knowledge goes, Eq. (\ref{etalimit}) is the only bound so far available for $R_{ab}R^{ab}$ theory alone.\\

A simpler version of our detailed analysis could be to compare the velocities of gravitational and electromagnetic waves, as done in most of the earlier cases \cite{gwspeed, jana}. An expression for the velocity of gravitational waves can be obtained from Eq. (\ref{eq:eff_metric}) and is given as $\sqrt{U/V}$. Since this varies with the redshift, one has to integrate over the path in order to find the full time delay, as has been done in our case. One can however make a rough estimation as follows. It is clear that the maximum deviation from GR occurs at $z=z_E$. One can therefore obtain a bound on $\alpha$ or $\eta$ by equating this value with the observed ($10^{-15}c$) deviation from the speed of light as $ \eta \,\Omega_m(1+z_E)^2 \approx10^{-15}$. This will give bounds of the same order as found by us. However doing the full integration is clearly the more precise approach and appropriate for sources at a high redshift. \\

\section{Discussions \& Conclusions}

The most straightforward extension of GR is the higher curvature gravity, where various higher curvature terms supplement the Einstein-Hilbert action functional. It is expected that the low-energy effective action of gravity will contain the higher curvature terms. These terms will be relevant at some length scale with new phenomenological effects. The path difference between gravitational waves and electromagnetic radiation is one of such effects which may reveal the scale of new gravitational physics.

In this work, we start with the simplest model for higher curvature gravity as in Eq. (\ref{lagra}), which can contribute to such a path difference. The path difference results in a delay between the detection of the gravitational waves and the associated electromagnetic radiation. The recently discovered source GW170817 and the electromagnetic counterpart GRB 170817A is a model system where we can study this effect. It was observed that the gravitational wave arrived about $1.7 $ s earlier than the electromagnetic wave. We work out an expression for the time delay, up to the first order in the higher curvature coupling. We also assume the existence of an intrinsic delay at the source due to various astrophysical effects. The final expression Eq. (\ref{DDNDOA}) gives the net observed delay. This equation is sufficiently general and can be applied to any source of gravitational waves with a known electromagnetic counterpart. 

To accurately estimate the higher curvature coupling from Eq. (\ref{DDNDOA}), we need to know the intrinsic delay at the source. At present, our understanding of the physics of gamma-ray burst cannot provide such information. As a result, we use various reasonable physical assumptions to find an upper limit on the coupling $\alpha$ in terms of the characteristic size of the Universe. Unlike tests based on other physical effects, which puts constrains on a combination of the couplings of $R_{ab}R^{ab}$ and $R^2$, our result can constraint the coefficient of the $R_{ab} R^{ab}$ term only, as other higher curvature term, which depends purely on the $\text{Ricci} \, \text{scalar}$, cannot cause any path difference, at least in the linear order of the coupling constants.\\

Though we can not accurately determine the intrinsic delay, the mechanisms which led to such a delay indicate that the GW should be emitted before the EM radiation. In our convention, this means $\delta t_E < 0$. If we also assume that $\alpha > 0$, then Eq. (\ref{etasource}) implies that for this system $ | \delta t_E (1+z_E) | \leq |\delta t_O |$ where the equality sign holds for the case of GR. This is the consequence of the fact that for positive values of the higher curvature coupling, the GW travels faster than EM radiation. So both the effects are working in the same direction.

The intrinsic delay, which results from the GRB physics, might also depend on the higher curvature coupling. One must then be able to expand it as a function of $\eta$. The terms $\delta t_E^{0},~\delta t_E^{1}...$, in Eq .(\ref{alphaexpansion}) are interpreted as coefficients of such an expansion. Though our final results do not take this into account, Eq .(\ref{alphaexpansion}) has this case incorporated. Then, the contribution to $\delta t_O$ from these higher-order terms will crucially depend on an intrinsic scale set by the GRB physics. It will be interesting to pursue a detailed investigation into this case, where there are two scales at play, the scale of the higher curvature term, set by $\alpha$, and the intrinsic scale, set by GRB physics.  

Given the parameters of the binary neutron star merger event GW170817, Eq. (\ref{DDNDOA}) gives a weak constraint on the coupling. But, the methodology developed in this work will result into much better bounds in the near future when we have more sources of simultaneous GW and EM emissions. Also, this formalism can be extended to any higher curvature theory of gravity for which the form of effective graviton metric is known.  At the same time, the result can be used to precisely estimate the intrinsic delay. This could be very useful to figure out the complex physics of gamma-ray bursts.\\ 

\section*{Acknowledgments}

We thank Sukanta Bose and Stefano Liberati for helpful discussions and suggestions. Research of SS is supported by the Department of Science and Technology, Government of India under the Fast Track Scheme for Young Scientists (YSS/2015/001346)). SS also thanks the hospitality of the Abdus Salam International Centre for Theoretical Physics (ICTP) where a part of this work is completed. AG is supported by SERB, government of India through the NPDF grant (PDF/2017/000533).

\end{document}